# Dynamical Coulomb Blockade as a Signature of the Sign-Reversing Cooper Pairing Potential

Chaofei Liu,[1,2] Pedro Portugal,[3] Yi Gao,[4] Jie Yang,[5] Xiuying Zhang,[6] Yanzhao Liu,[1] Tianheng Wei,[1] Wei Ren,[1] Jing Lu,[6,7,8,9] Christian Flindt,[3] and Jian Wang[1,7,10,†]

[1]*International Center for Quantum Materials, School of Physics, Peking University, Beijing 100871, China*
[2]*School of Physics and Wuhan National High Magnetic Field Center, Huazhong University of Science and Technology, Wuhan 430074, China*
[3]*Department of Applied Physics, Aalto University, Aalto 00076, Finland*
[4]*Center for Quantum Transport and Thermal Energy Science, Jiangsu Key Lab on Opto-Electronic Technology, School of Physics and Technology, Nanjing Normal University, Nanjing 210023, China*
[5]*Key Laboratory of Material Physics, Ministry of Education and School of Physics, Zhengzhou University, Zhengzhou 450001, China*
[6]*State Key Laboratory for Mesoscopic Physics and School of Physics, Peking University, Beijing 100871, China*
[7]*Collaborative Innovation Center of Quantum Matter, Beijing 100871, China*
[8]*Beijing Key Laboratory for Magnetoelectric Materials and Devices (BKL-MEMD), Beijing 100871, China*
[9]*Peking University Yangtze Delta Institute of Optoelectronics, Nantong 226000, China*
[10]*Hefei National Laboratory, Hefei 230088, China*
[†]jianwangphysics@pku.edu.cn

Coulomb blockade occurs for electrons tunneling into nanoislands because of the quantization of charge. Here, using spectroscopy measurements of nonmagnetic islands grown on a high-$T_c$ superconductor [one-unit-cell (1-UC) FeSe], we systematically investigate the dynamical Coulomb blockade (DCB), which is found to reflect the Cooper pairing potential in the superconducting substrate. The tunneling spectra are acquired on single-crystalline Pb nanoislands and show a clear suppression of the tunnel current around zero bias-voltage with a gap-like structure. The observed spectral gaps can be attributed to DCB based on our comprehensive investigations, including experiments with finely varying island sizes and calculations of the spectra using the $P(E)$ theory of DCB. Our detailed analysis suggests that the observed DCB can be related to the sign-reversing pairing potential in the 1-UC FeSe substrate below the islands. The sign reversal is furthermore revealed in a transition of the superconducting gap of FeSe from a U- to a V-like lineshape as the distance between neighboring doublet islands is decreased, indicating the presence of a nodal-like gap as expected for a sign-reversing superconductor. Our configuration of nonmagnetic nanoislands on a high-$T_c$ superconductor for spectroscopy measurements may serve as a local, spatially sensitive, and tunable probe for detecting the sign-reversing order parameter in unconventional superconductors.

## I. INTRODUCTION

One-unit-cell (1-UC) FeSe/SrTiO$_3$ has received increasing attention in recent years [1]. Compared to bulk FeSe, the significantly enhanced critical temperature (typically $T_c$ = 55–65 K) therein is unusual [2], since superconductivity should be suppressed by fluctuations in low-dimensional systems according to common believes [3, 4]. There is no Γ-hole pocket in 1-UC FeSe/SrTiO$_3$, thus the $s_\pm$-wave pairing based on electron–hole pocket nesting, normally expected in bulk iron-based superconductors [5-10], is unlikely. Theoretically, the pairing symmetry of 1-UC FeSe [11-17] allows both sign-preserving ($s_{++}$-wave) and sign-reversing pairings (e.g. incipient $s_\pm$-, extended $s_\pm$-, nodeless $d$-wave) [12, 18]. Therefore, phase-sensitive techniques are called upon to understand its pairing mechanism.

For small tunnel junctions with ultralow capacitances $C$, the combination of Coulomb charging effects and charge quantization strongly suppresses the tunnel current below the threshold voltage, $V_c=e/2C$, leading to Coulomb blockade [19, 20]. A typical scanning tunneling microscope (STM) setup, including the substrate-supported metallic nanocrystals [e.g. Fig. 1(a)], can be modeled as two junctions for describing the Coulomb blockade [21] [Fig. 1(b)]. Therein, the nanocrystal/substrate junction (numbered as junction #1), essentially considered as the external circuit where the tunneling barrier for STM is embedded, is characterized by the resistance $R_1$ and the capacitance $C_1$, in series with the STM tunnel junction ($R_2$, $C_2$) (junction #2). Interestingly, dynamical Coulomb blockade (DCB) occurs when single electrons tunneling across the barrier (junction #2) exchange energy with the local electromagnetic environment (incorporated into junction #1) through the emission or absorption of single energy quanta [Fig. 1(c)] [22, 23].

STM-based Coulomb-blockade phenomena have been observed in effective double junctions made of metallic clusters on surface-oxidized metallic substrates [19, 20, 24]. More recently, such experiments were extended to islands grown on purely metallic or semiconducting substrates (e.g., Cu [25], Al [26], HOPG [25], Si [25, 27], InAs [28]). Engineering the electrical contact between nanosized metallic crystals and their supporting substrates is important for designing future electronics. Still, the Coulomb-blockade effect has rarely been investigated for islands on high-$T_c$ superconductors. Moreover, the interplay between superconducting (SC) correlations and Coulomb interactions is still not well understood. In this work, metallic Pb nanocrystals with volumes $V$ well below the Anderson limit ($V_{Anderson} \simeq 100$ nm$^3$) [29] are grown on a high-$T_c$ substrate (1-UC FeSe/SrTiO$_3$) to explore the emergent effect of Coulomb blockade. The DCB, instead of the 'conventional'-type Coulomb blockade, is detected, and found suggestive of a sign-reversing pairing in the high-$T_c$ superconductor. Our results may stimulate the development of a phase-sensitive probe of Cooper pairing potentials based on Coulomb blockade.

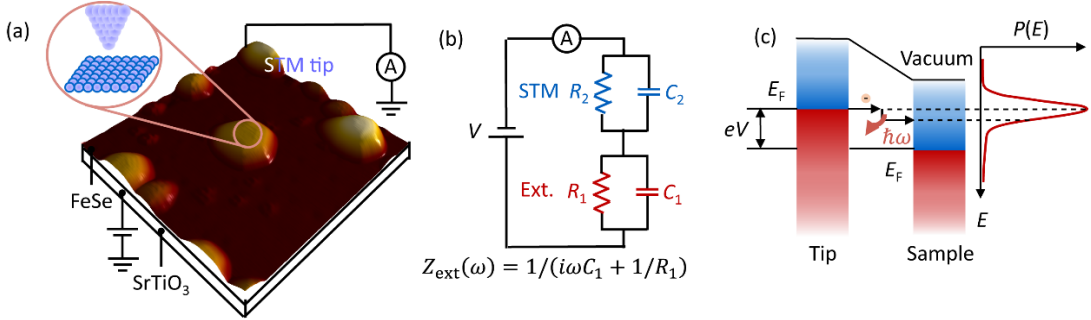

FIG. 1. Description of DCB in a two-junction system. (a) Schematic of the STM apparatus, showing electrons tunneling into the Pb nanocrystals grown on 1-UC FeSe/SrTiO$_3$. (b) Equivalent electrical circuit for the STM setup in (a). $Z_{ext}$ is the external impedance defined for junction #1. (c) Energy diagram of electrons tunneling in the DCB scenario. The tunneling electron interacts with the local electromagnetic environment by emitting or absorbing energy quanta of size $E = \hbar\omega$ with a probability given by the $P(E)$ function.

## II. EXPERIMENTS AND MODELING

### A. Sample growth and STM experiments

The experiments were performed in an ultrahigh-vacuum (5×10$^{-11}$–2×10$^{-10}$ mbar) STM system combined with a molecular beam epitaxy (MBE) chamber. The Nb-doped (0.7% wt.) SrTiO$_3$ substrate is prepared by a Se-flux-etching method [31]. The 1-UC FeSe film was epitaxially grown on Nb-doped SrTiO$_3$ following a well-established recipe [32]. By evaporating Pb from a standard Knudsen cell in the MBE chamber, high-purity Pb atoms were deposited on the substrate surface kept at room temperature at a rate of 0.17 monolayer (ML)/min. Via Volmer–Weber, i.e., island mode [33], the growth of Pb on 1-UC FeSe proceeded directly with the crystallization of individual nanoislands, without the formation of a 1–2 ML Pb wetting layer [34]. All the STM topographic images and tunneling spectra were measured at a temperature of 4.2 K unless specified. For the STM measurements, a bias voltage was equivalently applied to the sample. A mechanically sharpened polycrystalline PtIr tip was used throughout the experiments. The topographic images were obtained in a constant-current mode, with typical tunneling-junction setpoints: $V_s$ = 0.2–0.5 V, $I_t$ = 0.2–2.5 nA. The tunneling spectra were acquired at $V_s$ = 40 mV, $I_t$ = 2.5 nA using a standard lock-in technique with a bias modulation of 1 mV at 1.7699 kHz.

### B. Quantum-transport calculations

The 1-UC FeSe is firstly fully optimized with the lattice parameter converged at $a_0=b_0$=3.70 Å. The 1-UC FeSe and Pb metal interface system is then constructed by putting 1-UC FeSe on one side of the Pb metal surface, and the interlayer distance between the two surface is initially set to be about 3.2 Å. The Pb metal surface is simulated with a slab of six Pb atom layers along the <111> direction. The lattices of the Pb slab are changed to match a 3×2 1-UC FeSe supercell with a mismatch of 4.6%, as the Pb island is later grown onto 1-UC FeSe. The lattice parameters of the interface system are kept during the geometrical optimization process. A vacuum buffer space is set to be at least 12 Å to avoid spurious interactions.

The geometry optimization and electronic property calculation are performed within the Vienna *ab initio* simulation package (VASP). The plane-wave basis set with the cut-off energy of 600 eV and projector augmented wave (PAW) pseudopotential is employed. The exchange-correlation functional is the generalized gradient approximation (GGA) with the Perdew−Burke−Ernzerhof (PBE) parametrization. A fine $k$-mesh density of 0.02 Å$^{-1}$ under the Monkhorst−Pack method is sampled in the Brillouin zone. The atoms are relaxed until the residual force is less than 0.01 eV·Å$^{-1}$/atom and the total energies converged to less than $1\times10^{-6}$ eV/atom.

A two-probe device configuration is built to simulate the transportation process between the 1-UC FeSe and Pb island. The 1-UC FeSe and Pb interface system is used as channel, and the metal Pb and 1-UC FeSe are used as left and right electrode, respectively. The building of the device and transport property calculations are implemented in the Atomistix ToolKit 2020 package, and use a nonequilibrium Green's function (NEGF) method coupled with density functional theory. The transmission coefficient $\tau_{k_\parallel}(E)$ can be obtained in the irreducible Brillouin zone (IBZ) as

$$\tau_{k_\parallel}(E) = Tr\left[\Gamma^l_{k_\parallel}(E) G_{k_\parallel}(E) \Gamma^r_{k_\parallel}(E) G^\dagger_{k_\parallel}(E)\right]$$

where $G_{k_\parallel}(E)$ and $G^\dagger_{k_\parallel}(E)$ are the retarded and advanced Green's functions, respectively. $\Gamma^{l(r)}_{k_\parallel}(E) = i\left(\sum_{l(r),k_\parallel}(E) - \sum^\dagger_{l(r),k_\parallel}(E)\right)$ represents the level broadening originating from the left and right electrodes in the form of the self-energy $\sum_{l(r),k_\parallel}(E)$, which reflects the influence of the electrodes on the scattering region. $k_\parallel$ is a reciprocal lattice vector pointing along a surface-parallel direction (orthogonal to the transmission direction) in the IBZ. Double-ζ plus polarization (DZP) basis set is employed, the kinetic-energy cutoff is of 120 Hartree, and the temperature is set at 300 K. The $X$ and $Y$ directions of the device take Neumann and periodic boundary condition, respectively. The transport $Z$ direction takes a Dirichlet-type boundary condition.

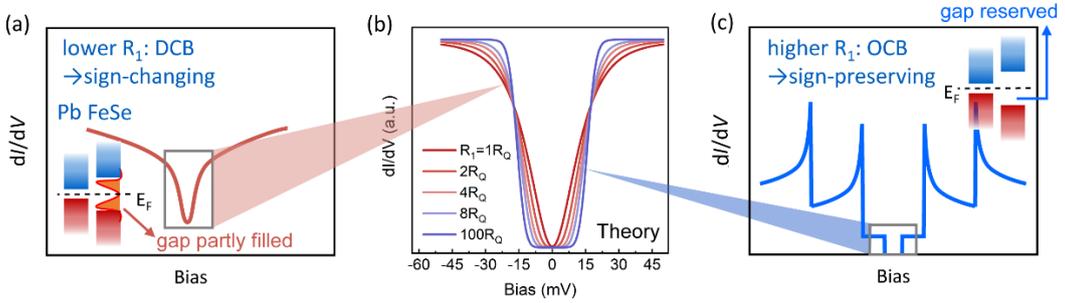

FIG. 2. DCB for probing the Cooper pairing. (a,c) Schematic illustrations of DCB and OCB tunneling spectra, expected for a nonmagnetic nanoisland/superconductor system. The OCB spectrum is accompanied by a Coulomb staircase, corresponding to whenever an extra electron is accommodated in the central electrode (i.e., the nanoisland, or nanocrystal in our work). (b) Examples of d$I$/d$V$ spectra for different $R_1$ calculated using the $P(E)$ theory of DCB [25]. From (a)-(c), the two kinds of Coulomb-blockade phenomena in (a) and (c) exhibit different low-energy gaps (boxes), highlighting the low- and high-$R_1$ contacts, respectively, based on the comparison with $R_1=R_Q$ and $=100R_Q$ spectra in (b). A sign-changing (-preserving) pairing corresponds to the lower (higher) $R_1$ due to the presence (absence) of intragap states [insets of (a) and (c)].

## III. COULOMB BLOCKADE AND COOPER PAIRING

### A. Coulomb blockade as a possible probe of the Cooper pairing potential

For Coulomb blockade, due to the quantized electronic-charging effects, electrons only tunnel when the applied voltage bias $V$ can overcome the charging energy, i.e. $e|V|\geq E_c=e^2/2C$, yielding the threshold voltage $V_c=e/2C$. Below $|V_c|$, because of the blocked electron accumulation by Coulomb repulsion, tunneling is strongly suppressed. Benefiting from technological advances in nano-fabrication techniques and STM, the Coulomb-blockade effect has been explored in elaborate planar or point-contact junctions with ultralow capacitances [35]. These investigations are not only important for understanding the physics of small conductors in low dimensions [19], but also pave the way for a range of electronic applications, such as current standards, electrometry, fast-switching devices, and single-electron memories [36-38]. Two preconditions are crucial for Coulomb blockade [19, 21]: i) low temperature, to ensure the charging energy $E_c$ is higher than the thermal energy $k_B T$; ii) sufficiently large tunnel resistance $R$ compared with the quantum resistance, $R_Q = h/e^2 \simeq 25.8$ kΩ, to suppress the smearing by quantum fluctuations. For the STM-based double junctions, the operating temperature of 4.2 K is low enough to neglect thermal effects, and the tip–island junction resistance $R_2$ (typically 1 MΩ–1 GΩ) is always significantly larger than $R_Q$, guaranteeing that both conditions are satisfied.

Based on the $P(E)$ theory of DCB [23, 25, 39-42], we expect that the low-energy tunneling spectra (d$I$/d$V$ vs. $V$) evolve from a soft to a hard gap, when the Pb/FeSe contact resistance is increased from $R_1 \simeq R_Q$ to $\simeq 100 R_Q$ [Fig. 2(b)]. Essentially, the soft gap for the low-$R_1$ junction is typical of a DCB signal [Fig. 2(a)] [22, 23]. By contrast, the hard gap for the high-$R_1$ junction—normally flanked by the Coulomb staircase, falls within the well-known picture of orthodox Coulomb blockade (OCB) [Fig. 2(c)] [39]. Notably, the Coulomb-gapped spectral segment of OCB near zero energy remains robustly captured by the $P(E)$ theory in the high-$R_1$ limit [Fig. 2(b)].

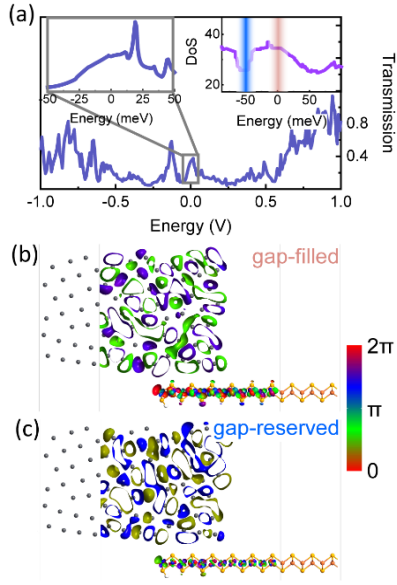

FIG. 3. Transmission spectrum. (a) Calculated full-energy transmission spectrum within [−1,1] eV for a Pb/FeSe structure. Insets, low-energy transmission and DoS spectra. (b,c) Wavefunctions, imprinted on the Pb/FeSe junction, of the dominating eigen-transmission channels for eigenenergies at $E_F$ and −54 meV, which are near the gap-filled and gap-reserved limits, respectively.

To demonstrate the possible relation between Coulomb blockade and Cooper pairing, the high-$T_c$ 1-UC FeSe is chosen because of its large, tunnel-barrier-type SC gap (10–20 meV [12]) for the contact to metallic Pb islands. Based on the selection rules for Bogoliubov quasiparticles [43], as the non-magnetic scatterer [44], Pb selectively depairs the sign-reversing pairings (e.g. $s_\pm$-, $d$-wave) with induced intragap excitations, but keeps the sign-preserving pairing ($s$-wave). Previous studies have established that the scatterers, either in the form of single adatoms or nm-scale islands [45-47], can equivalently induce sharp in-gap states [45, 47] or gap filling and narrowing [46], when superconductivity is locally suppressed. If sign-reversing (or sign-changing) $s_\pm$- or $d$-wave pairing dominates in FeSe, the SC gap will be filled with Pb-island-induced intragap states [inset, Fig. 2(a); termed 'gap-filled' case], lowering the barrier. Thus, the electrical Pb nanoisland/FeSe substrate contact could be as low as $R_1 \simeq R_Q$ [25], which would give DCB tunneling spectra [Fig. 2(a)]. Yet, if a sign-preserving $s$-wave pairing dominates, the FeSe substrate should remain fully gapped [inset, Fig. 2(c); 'gap-reserved' case] with a large barrier resistance to the Pb island, leading to $R_1 \gg R_Q$ [28, 48]. Then, the spectrum for tunneling into the Pb island should show OCB accompanied by Coulomb-staircase features [Fig. 2(c)]. Taken together, by investigating the electrical contact between nonmagnetic islands and the SC substrate through the Coulomb-blockade phenomena, we can obtain possible insights into the pairing scenario for the SC substrate underneath.

## B. Quantum transport simulations

To quantitatively support the conjecture discussed above, concrete values of $R_1$ are evaluated by calculating the transmission spectrum $\tau(E)$ of an experiment-based Pb/FeSe structure. The effective contact area $S$ is fixed to the optimized size ($S_{calc} \sim 0.82$ nm$^2$) of a 3×2 FeSe supercell. In the 'normal' (i.e. non-SC) state of Pb/1-UC FeSe, the density of states (DoS) shows metallic and partly gapped behaviors near the Fermi level $E_F$ and at −54 meV, respectively [right inset, Fig. 3(a)], which are adopted to simulate separately the two situations towards gap-filled and gap-reserved limits. In practice, $R_1$ may be affected by many factors, e.g., the imperfect interface contact because of unknown buried impurities/defects or 'bubbles'. However, all of these factors influence the resistance by essentially changing the transmission. A more complicated microscopic description might include these factors; yet, in our simulations, they have been effectively considered by properly choosing the gap filling.

Figure 3(a) shows the calculated full-energy transmission in the main panel, while the left inset presents the zoom-in of the low-energy part with the energy range as in the measurements. Near the gap-filled limit corresponding to the sign-reversing pairing case, i.e. at $E_F$, we find $\tau$=0.34, yielding $R_1=R_Q/\tau \simeq 2.9 R_Q$, meeting the condition of $R_1 \approx R_Q$ for the occurrence of DCB [Fig. 2(a)]; similarly, near the gap-reserved limit for the sign-preserving pairing case, we have $\tau$=0.085, yielding $R_1=R_Q/\tau \simeq 11.8 R_Q$, indicating that $R_1 \gg R_Q$ is satisfied for the occurrence of OCB [Fig. 2(c)]. Consistently, the dominating transmission channel shows decreased tunneling probability in the gap-reserved case, as reflected in its suppressed wavefunctions transmitted into the FeSe layer [Fig. 3(c)] compared to the gap-filled situation [Fig. 3(b)].

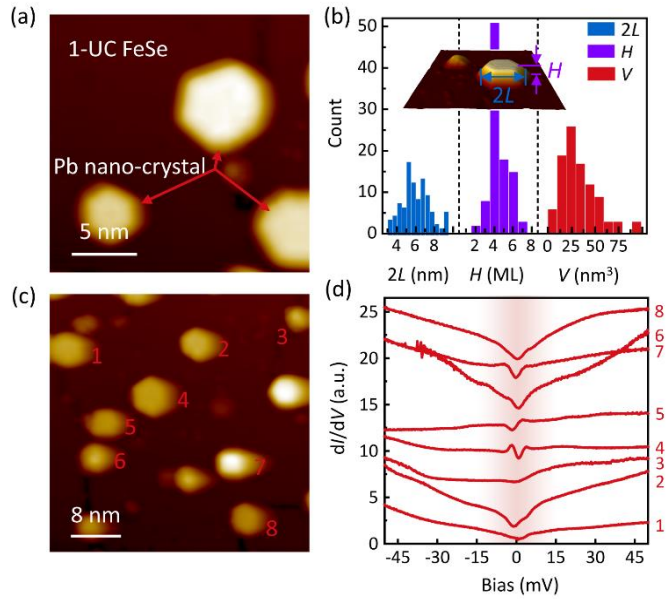

FIG. 4. Detection of DCB. (a) STM image of Pb nanocrystals on 1-UC FeSe/SrTiO$_3$(001). (b) Distribution of the lateral size $2L$, height $H$ (inset), and volume $V$ (assuming $V = \frac{3\sqrt{3}}{2}L^2 H$) of Pb nanocrystals. (c,d) Numbered Pb nanocrystals and associated tunneling spectra.

## IV. RESULTS

### A. Probing the sign-reversing pairing via DCB

Our metallic Pb nanocrystals were prepared *in situ* by depositing Pb on a high-quality 1-UC FeSe/SrTiO$_3$ (Fig. S1 [49]) in an MBE system. Single-crystalline Pb nanocrystals, or nanoislands, formed with straight edges, flat tops, and hexagonal shapes, indicating an exposed [111]-oriented structure [Fig. 4(a); Fig. S2 [49]]. Compared with the Pb nanocrystals previously grown on Si(111) [25, 27, 50], SiC [51], or metallic substrates (Cu, Ag [25]), our Pb islands show particularly well-defined hexagonal shapes, illustrating their high crystalline quality. The Pb nanoislands have lateral sizes of $2L$=3–9 nm ($L$, side length of hexagon), and thicknesses of $H$=2–7 ML (1 ML=0.286 nm) [Fig. 4(b)].

Using $V = \frac{3\sqrt{3}}{2} L^2 H$ for a hexagonal prism, we found volumes in the range $V$=4–100 nm$^3$, corresponding to $0.04 V_{\text{Anderson}}$–$1 V_{\text{Anderson}}$. Thus, we obtained a high-$T_c$ substrate supporting well-isolated nonmagnetic nanocrystals with volumes below the Anderson limit.

To examine which type of Coulomb blockade is present, we conducted systematic tunneling-spectroscopy measurements on the FeSe-supported Pb nanocrystals. In the d$I$/d$V$ spectra, a soft gap of ~5 meV, which is usually uniform in space within each nanocrystal (Fig. S3 [49]), appears around zero bias with a strong tunnel-current suppression [Figs. 4(c) and 4(d)]. Evidently, despite the differences between individual Pb nanocrystals, the soft tunneling gap is reproducibly detected. Because of the difference of the nanocrystals in size and their local environments, including the nano-scale inhomogeneity of disorder and defects beneath the nano-islands, the soft gap is sometimes accompanied by irregular spectral features outside the gap.

We also investigated the spectral properties of the FeSe substrate in the immediate vicinity of the Pb nanocrystals. The results show that the fully gapped, spatially homogeneous SC spectrum in the FeSe substrate remains nearly unchanged independently of the distance to the Pb islands [Figs. 5(a) and 5(b)]. As the STM tip crosses the Pb-nanocrystal edges, the spectra abruptly turn into soft gaps with a finite zero-bias differential conductance (d$I$/d$V$) [Figs. 5(b) and 5(d)]. At large bias voltages, e.g., at −30mV, the spectra with and without the Pb islands are also strikingly different, similarly showing a sharp change [Fig. 5(c)]. The full spectral structure in space is clearly seen in the zero-bias d$I$/d$V$ map for the Pb nanocrystal [Fig. 5(e)]. Evidently, the finite island conductance (light red) sharply contrasts with the vanishing FeSe conductance (gray). The well-defined conductance hexagon, corresponding to the Pb-nanocrystal profile, illustrates how the spectra change abruptly. The highly localized nature of the soft-gap spectra on the nanocrystal region highlights its prospect of probing the possible SC pairing with nanoscale sensitivity.

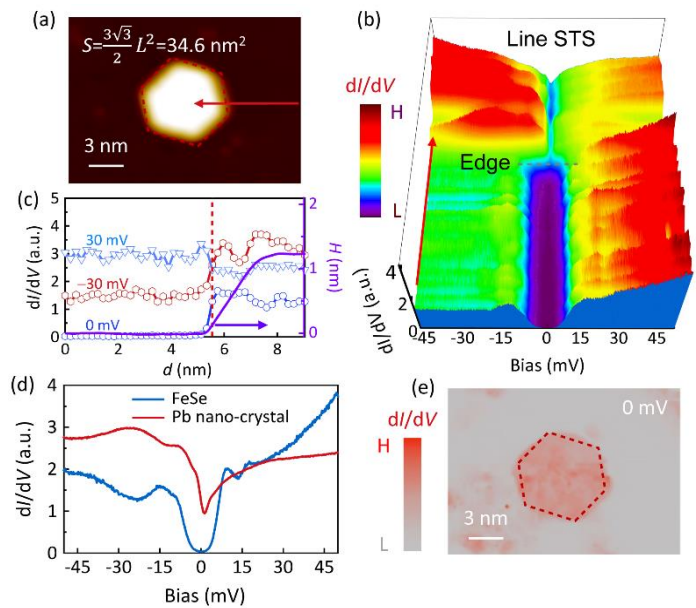

FIG. 5. High spatial sensitivity of soft-gap signal. (a,b) STM image of a Pb nanocrystal, and tunneling spectra measured along the arrow-indicated path therein. See Figs. S4–S6 [49] for edge-crossing spectral-evolution data likewise. (c) Height $H$ profiles and related d$I$/d$V$ at selected biases (0, −30 and 30 mV) along the same path, where $d$ is the distance. (d) Tunneling spectra taken on the Pb nanocrystal and directly on the FeSe substrate, showing DCB and SC gaps, respectively. (e) Zero-bias d$I$/d$V$ map measured for the Pb nanocrystal in (a). The dashed hexagons indicate the Pb-nanocrystal edges. See Part SIII [49] for reproduced DCB gaps with high spatial sensitivity.

Given the nano-scale size of the Pb islands, together with the preconditions for Coulomb blockade satisfied in our STM apparatus and the spectral lineshape as predicted by the theory of DCB, we associate the soft gap to DCB effects. Here, the DCB occurs in double tunnel junctions with relatively low electrical contact ($R_1$) and high capacitances ($C_1$) for junction #1 [21, 25, 26, 52]—in our experiments, formed between the Pb nanoislands and the FeSe substrate. Based on the above arguments, the observation of DCB is a direct manifestation of the low-$R_1$ contact. Evidently, the small $R_1$ cannot be explained by a structurally imperfect contact that instead tends to increase the interface barrier, and is most likely attributed to the gap-filling intragap excitations induced by the ensemble-type nonmagnetic scatterers of Pb nanoislands. One may further argue that even for the sign-preserved pairing, the $R_1$ resistance can be low enough to cause DCB. However, in the sign-preserved case, we have adopted a gap-filling factor of ~0.75, defined as $v = \frac{dI}{dV}_{\text{gap center}} / \frac{dI}{dV}_{\text{gap edge}}$, for our simulations [Fig. 3(a)], which is overestimated even if one includes contributions from the possible in-gap excitations due to Andreev bound states, thermally excited intra-gap quasiparticles, etc. Yet, the resulting value of $R_1 \simeq 11.8 R_Q$ still exceeds $R_Q$ by far, meaning that DCB is less plausible in the sign-preserving situation. Since the Andreev bound states exist independently of the sign structure, the low $R_1$ for DCB case could be a combined effect of the sign reversal and the Andreev states. Thus, the result is expected to contain information about a sign-reversing pairing of 1-UC FeSe substrate, which is also hinted in earlier observations [16, 44, 53] and the cooperative-pairing conjecture [54].

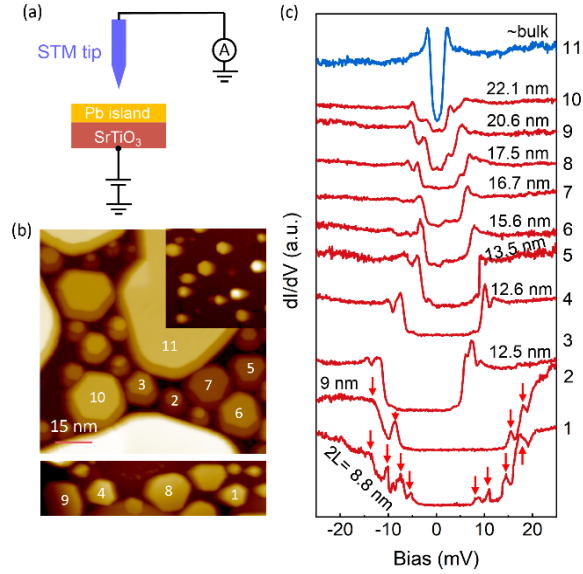

FIG. 6. OCB detected on the Pb nanoislands directly grown on $SrTiO_3$. (a) Schematic of STM-measured Pb nanoislands/$SrTiO_3$. (b) STM image of the Pb nanocrystals/$SrTiO_3$. Inset shows the image of Pb nanocrystals/1-UC FeSe sharing the same scale bar as the main panel for a direct size comparison of islands on different substrates. (c) Tunneling spectra taken on the $SrTiO_3$-supported Pb nanocrystals [as numbered in (b)] with the size comparable, or a little larger than those on FeSe (except #11, a control spectrum for bulk-like Pb island). The spectra (vertically offset) are ordered by island size ($2L$) as marked. The discrete peaks—the signatures of Coulomb blockade, are marked by arrows for the bottom spectra as an example.

### B. Control experiments

To check the role of the contact resistance ($R_1$) on Coulomb blockade, we performed control experiments with Pb nanoislands grown directly on $SrTiO_3$ (i.e. without the FeSe layer) [Figs. 6(a) and 6(b)], a dilute-doping semiconductor. $SrTiO_3$, with a large bandgap (2–3 eV [55, 56]) supposedly serves as a large barrier. Thus, the contact resistance should be larger than the case of partly intragap-filled FeSe. Tunneling $dI/dV$ spectra are taken at $SrTiO_3$-supported Pb nanocrystals [Fig. 6(c)] with comparable, or slightly larger size relative to those showing DCB on FeSe ($2L \simeq 3$–9 nm). The results include that, i) the $dI/dV$ spectrum for a small island with $2L = 8.8$ nm shows discrete Coulomb peaks outside a central hard Coulomb gap near zero voltage; ii) as the size increases (overall $2L \lesssim 22$ nm here), the spectrum evolution presents a trend to a lineshape showing only the central hard gap flanked by initially developed one or two pairs of Coulomb peaks. These observations in both situations represent the signatures of OCB. Predominately, the gap is contributed by the Coulomb blockade-gapped feature mainly arising due to the island–substrate junction. Component from an interaction effect-induced correlation gap is additionally superposed, which usually appears in dimension-reduced, ultra-small sized systems due to suppressed Coulomb screening therein. The contribution from the correlated gap is expected to decrease when the island size gets larger [57-60], as observed here. The bias-asymmetric lineshape for some spectra is attributed to the residual fractional charge $Q_0$ in the nanocrystals [39]. In contrast, the islands with size comparable with, or larger than the bulk SC coherence length $\xi$ ($\xi_{Pb}$ = 83 nm), still show well-defined SC spectra as expected (Fig. S9), indicating that the OCB signals are indeed reliable. The OCB found for tunneling across a large-$R_1$ contact barrier in turn indicates that DCB occurs in the Pb/FeSe junctions with a lower-contact nature.

### C. Further proof of DCB via the island-size dependence

Independent evidence for DCB is provided by the dependence of the soft gap on the nanocrystal's surface area. For the double junction with nanocrystal surface area $S$, the capacitance reads $C = \varepsilon_r\varepsilon_0 S/\delta$ ($\delta$, effective junction thickness). Accordingly, the Coulomb-blockade threshold, $V_c = e/2C$, suggests that with increasing $S$, the DCB gap size should decrease [20]. In Fig. 7(a), we present $dI/dV$ spectra for more than 50 nanocrystals with increasing $S$. These results show that, despite the spectral fluctuations, the gap width (blue region) generally becomes smaller as the island size $S$ increases, which is consistent with the DCB description. To further provide quantitative evidence for DCB, we fitted the tunneling spectra using the $P(E)$ theory of DCB with $R_1$ and $C_1$ as fitting parameters [25]. The fitted spectra for the measured data capture the statistical dependence of the Coulomb gaps on the island sizes [Fig. 7(b)] and the detailed spectral lineshapes observed experimentally [Fig. 7(c)]. Note that some uncertain factors, such as a change of the STM tip's atomic configuration or unknown impurities below the islands, may partly affect the measured spectra and make them deviate from the lineshape predicted for DCB. The fittings also yield reasonable values of the resistances and the capacitances for electrical contacts [Figs. 7(d) and 7(e)]. Specifically, the obtained average capacitance between the Pb islands and FeSe (i.e. $\overline{C_1}$=14.2 aF) dominates over the capacitance between the islands and the STM tip ($C_2 \lesssim 1$ aF), which is a typical phenonmenon of STM-based DCB [25], and the resistance $R_1$ spans a range of ~0.05$R_Q$–2$R_Q$ corresponding to the low-impedance regime for DCB [61]. As expected, $C_1$ and $R_1$ scale positively with $S$ and $1/S$ in statistics, respectively, further supporting the DCB description based on $P(E)$ theory. Furthermore, the extrapolated $R_1$ at $S_{calc}$~0.82 nm$^2$ following the linear fit of measured $R_1$ vs. $1/S$ data [Fig. 7(e)] turns out to be 2.45$R_Q$, in agreement with the calculated $R_1$=2.9$R_Q$ [Fig. 3(a)].

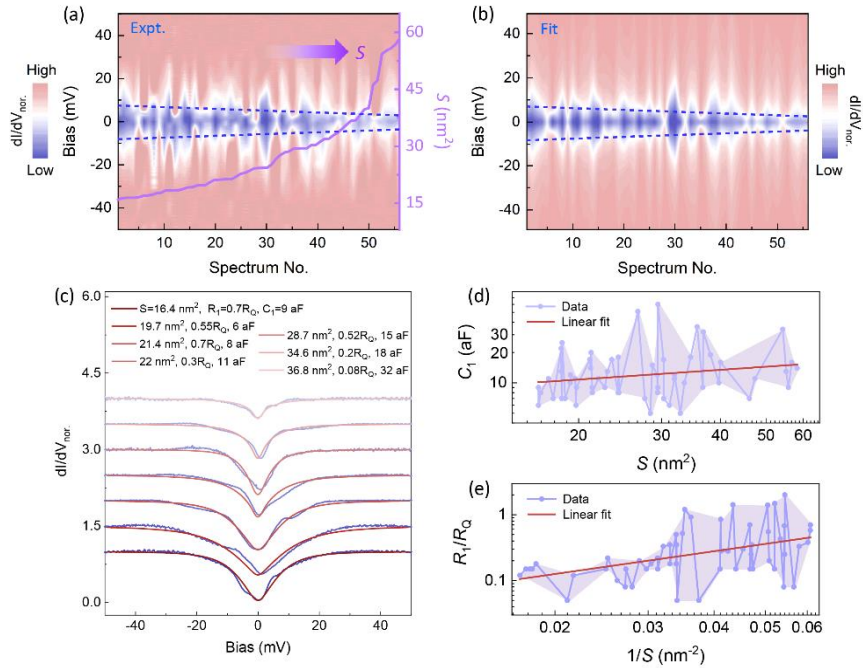

FIG. 7. Island-area dependence of the DCB gap. (a,b) Measured and associated $P(E)$-theory-fitted [25] DCB spectra for 56 Pb islands on 1-UC FeSe. The data are normalized (Fig. S10 [49]), and ordered successively with assigned 'spectrum No.' according to the monotonically increasing island surface area, $S$ ($=\frac{3\sqrt{3}}{2}L^2$). $S$ vs. spectrum No. is shown with the purple curve in (a). (c) Comparison of experiment and fit for a few selected tunneling spectra (blue lines, experiment; red lines, fit). (d,e) Fitting parameters, $C_1$ vs. $S$ and $R_1/R_Q$ vs. $1/S$.

## D. Sign reversal revealed by a nodal transition

The DCB-revealed gap-sign reversal, if any, can also in principle be observed from the transition of the SC spectral lineshape of FeSe substrate. Theoretically, the sign-reversing pairing would yield a V-shaped $dI/dV$ spectrum [65] when nodal lines (i.e. the collection of zero-gap criticality) cross the Fermi surface [66], otherwise the SC lineshape remains nodeless and U-gapped. In our experiment, the spectra taken along an 'open' trajectory, e.g. the red-arrow-marked one in Fig. 8(a), show a reserved U-like lineshape as normally expected for 1-UC FeSe [Fig. 8(d)]. However, for the trajectory between two adjacent Pb nanocrystals, the spectra become V-shaped with enhanced low-energy quasiparticle excitations [Figs. 8(b) and 8(c)]. Based on further statistics over 24 sets of doublet Pb islands, as the nanocrystal-to-nanocrystal distance $2d_\Delta$ is decreased, the U-gapped SC spectra collected in-between indeed gradually evolve into V-like shapes [Fig. 8(e)]. To quantify the evolution, we fit the low-energy ($|V|\lesssim 10$ mV) spectral parts using an empirical power law, $dI/dV \propto |V|^\alpha$ [inset, Fig. 8(f)], as adopted previously for bulk FeSe [67]. The extracted $d_\Delta$-dependent exponent $\alpha$ shows a transition to hard-gap opening at $d_\Delta \gtrsim 3.2$ nm [Fig. 8(f)]. Each $\alpha$ value in Fig. 8(f) is extracted from a single spectrum at the 'middle point' of the doublet islands and is sufficiently representative, considering the $\alpha$'s spatial variation exerts negligible influence on the judgement of U or V lineshapes (Fig. S14 [49]). Such a 'nodal transition' suggests that a nodal-gap structure can exist in 1-UC FeSe/SrTiO$_3$ in specific situations, which supports the sign-reversing gap scenario as implied from our DCB results.

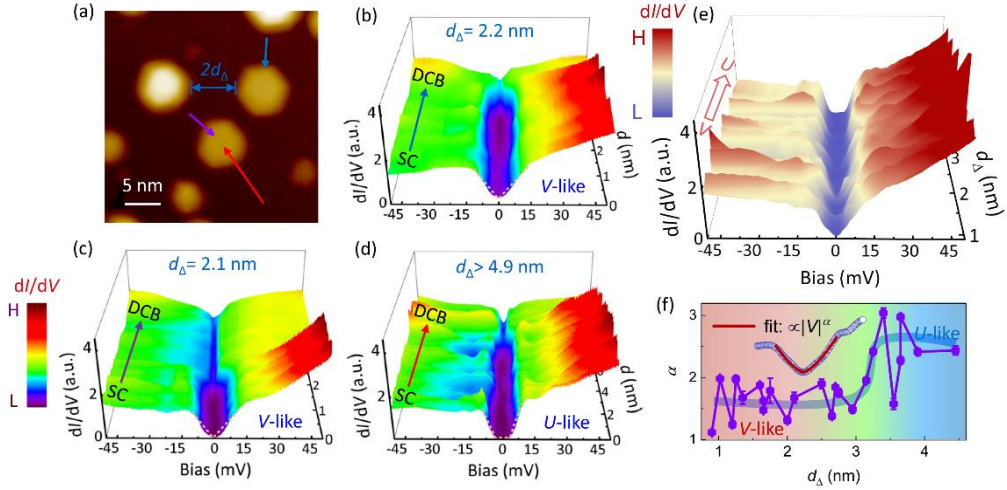

FIG. 8. Effect of Pb nanocrystals on the SC-gap structure of 1-UC FeSe. (a) STM image of Pb nanoislands. $d_\Delta$, half inter-island distance. (b–d) Tunneling spectra taken along the arrows in (a). For more examples of similar phenomena, see Fig. S11 [49]. (e) Lineshape evolution of SC spectra taken at the middle points (defined in Fig. S12 [49]) of 24 sets of doublet islands as a function of $d_\Delta$ (see Fig. S13 [49] for the raw data). (f) Summary of $d_\Delta$-dependent spectral lineshape, quantified by the exponent $\alpha$ as determined from the fit exemplified in the top-left inset.

## E. Simulations of the spectral U–V transition

Essentially, the Pb nanoisland can be regarded as a finite-size 'impurity' yet with stronger scattering potential given its nature as an ensemble of scatterers [43, 68]. To understand the observed U–V gap transition more rigorously, we calculated the DoS spectra at the middle site of two nonmagnetic impurities with different distances $N_\Delta$ [Fig. 9(a)] under various pairing scenarios. Calculations of band structures and pairing-gap function of 1-UC FeSe/SrTiO$_3$ are based on a previously proposed 2D tight-binding model [69, 70] (see Methods in Ref. [44]). The impurity Hamiltonian can be written as $H_{\text{imp}} = \sum_J \sum_{\alpha,\beta=1}^{2} \left(V_{\text{p},\alpha\beta}^J + V_{\text{m},\alpha\beta}^J\right) c_{R_J A \alpha \uparrow}^\dagger c_{R_J A \beta \uparrow} + \left(V_{\text{p},\alpha\beta}^J - V_{\text{m},\alpha\beta}^J\right) c_{R_J A \alpha \downarrow}^\dagger c_{R_J A \beta \downarrow}$. Here, $J$ (=1, 2) is the index of impurities located at the Fe atom of sublattice A in the 2-Fe unit cell, with the impurity-site coordinates denoted as $\boldsymbol{R}_J$, and $V_{\text{p},\alpha\beta}^J$ is the strength of the potential (nonmagnetic) scattering [intra- ($\alpha = \beta$) or interorbital ($\alpha \neq \beta$)] for the impurity (assuming $V_{\text{p},\alpha\beta}^1 = V_{\text{p},\alpha\beta}^2$).

The simulated 'middle-site' spectrum, exemplified under nodeless *d*-wave pairing, appears gap-filled by impurity bound states, and as observed in experiments, turns towards V-shaped (although not ideally) when significantly decreasing $N_\Delta$ [Fig. 9(b)]. For other pairing scenarios, at sufficiently small $N_\Delta=2$, the simulated spectrum remains U-gapped under isotropic *s*- and hidden $s_\pm$-wave pairings, while it becomes V-like under extended $s_\pm$-wave pairing (Part SVII [49]). These observations coincide with the fact that, among the four pairing candidates, sign reversal near $E_F$ occurs only in extended $s_\pm$- and nodeless *d*-wave pairings, further supporting the spectral U–V crossover taken as the signature of sign-reversing pairing. Regarding the possible physics picture of the U–V-shaped transition, the V gap appearing only for sufficiently adjacent doublet nanoislands suggests that the scattering potential enhanced at the exposed FeSe region in-between is responsible. Specifically, such enhancement can be attributed to the superposition of the potential provided by individual members of the doublet islands, which modifies the Fermi level ($E_F$) to an extent such that $E_F$ crosses the nodal line to trigger the nodal transition (Fig. 10). Generally speaking, the variation of Fermi energy can be revealed experimentally via tracing the energy shift of a characteristic state. Nevertheless, for strongly correlated SC systems, with the change of Fermi energy, the nonnegligible correlation effect will inevitably modify the spectrum lineshape, making the shift of a characteristic state be hardly identifiable in experiments.

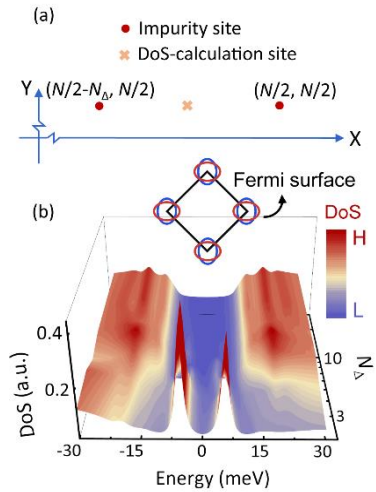

FIG. 9. Spectral simulations at the inter-impurity site. (a) Schematic showing the DoS-calculation site in the middle of two nonmagnetic impurities at the Fe atom of sublattice A, with the location sites $\boldsymbol{R}_J = (N/2−N_\Delta, N/2)$ and $\boldsymbol{R}_{J'} = (N/2, N/2)$. (b) Calculated 'middle-site' DoS in nodeless *d*-wave pairing scenario for various inter-impurity distances $N_\Delta$. Inset, Fermi-surface topology in folded Brillouin zone. $V_{p,11} = V_{p,22} = 1700$ meV, $V_{p,12} = V_{p,21} = 1300$ meV.

## V. DISCUSSIONS

### A. Possible origins of the soft gap in Pb nanocrystals

Given a critical temperature of $T_c = 7.2$ K for bulk Pb, our soft gap detected at 4.2 K resembles a SC gap, either intrinsic to Pb nanocrystals or induced by the proximity effect from FeSe. However, this possibility can be excluded, because:

i) The Pb islands on 1-UC FeSe here are below the Anderson limit, where superconductivity is expected to be quenched [28, 71, 72]. The volume range ($0.04V_\text{Anderson}$–$1V_\text{Anderson}$) spanned by the islands includes the borderline sizes of the Anderson limit and may support the possibility that Anderson criterion cannot be applied. Yet, the volume statistics [Fig. 4(b)] shows that the majority (~81%) of the islands are well below the Anderson limit ($\leq 0.5V_\text{Anderson}$) with quenched superconductivity for sure, while those with volume of $0.6V_\text{Anderson}$–$1V_\text{Anderson}$ that may be taken as 'borderline cases' only take up a small proportion of ~10%. Since no abrupt change of the spectral lineshape is identifiable, both the majority- and the minority-sized Pb islands are believed to be of the same origin with the quench of superconductivity.

ii) The Pb-islands' d*I*/d*V* spectra mostly show no SC coherence peaks flanking the gap. In comparison, the d*I*/d*V*

lineshape of the spectra observed in Pb nanoislands grown on FeSe is significantly different from that of SC spectra taken on much larger-sized Pb islands with well preserved superconductivity (Fig. 11).

iii) The soft gap reproducibly persists above the $T_c$ of bulk Pb (Fig. 12).

iv) The gap suppression with increasing size [Fig. 7(a)] is opposite to the behavior of superconductivity.

v) Upon crossing the Pb island/FeSe edge, the spectral lineshape shows a sharp transition [Fig. 5(b)], rather than a continuous evolution expected for proximity-induced superconductivity.

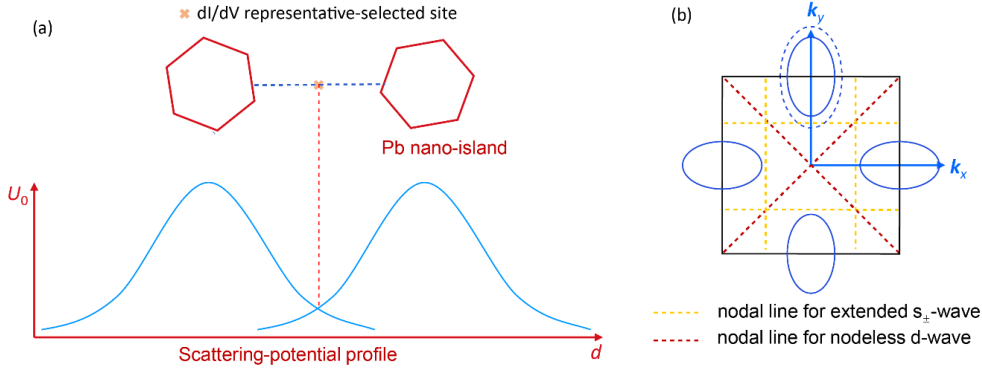

FIG. 10. Scenario of the nodal transition. (a) The profile of the scattering potential for the doublet islands, which is enhanced near the middle-site region compared to that of a single island. (b) Fermi surface of 1-UC FeSe shown in the unfolded Brillouin zone as an example. The yellow and red dotted curves denote the nodal lines for extended $s_\pm$- (gap function: $\Delta=\Delta_0|\cos k_x \cos k_y|$) and nodeless $d$-wave pairing ($\Delta=\Delta_0|(\cos k_x - \cos k_y)/2|$), respectively. The enhanced scattering potential between the doublet islands—especially near the middle-site region, when sufficiently strong, will increase the Fermi level to a critical point, where the enlarged electron-type Fermi pockets (dashed ellipse) intersects with the nodal lines, i.e., the nodal transition occurs.

Other possible explanations of the soft gap could be electron–electron interactions and reduced electron–phonon scattering. The soft gap induced by the former (electron–electron interactions) is predicted for weakly disordered metals [73], e.g. amorphous alloys and granular metals, which are clearly different from our Pb nanocrystals with crystalline quality. The gap related to the latter mechanism is 'mediated' by (or accompanied by) quantum well states (QWSs) of Pb islands [74], and appears due to the improved interference of the elongated quasiparticle lifetime within $\pm E_D$ (Debye energy) when the Fermi level is in between the QWSs peaks. No signatures of QWSs are detected in our nano-islands, suggesting that reduced electron–phonon scattering is unlikely. Even though they are inapplicable to our situation, future field-, temperature- or substrate-dependent investigations may help rule them out decisively.

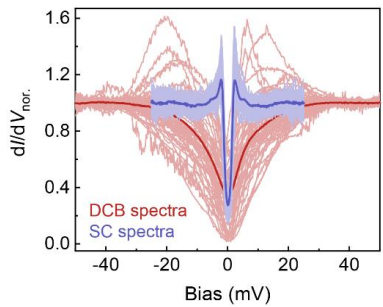

FIG. 11. Comparison between DCB and SC spectra. Normalized $dI/dV$ spectra obtained on smaller- [$2L\approx$3–9 nm, Fig. 4(b); red] and larger-sized ($2L\gtrsim$50 nm; light blue) Pb islands grown on FeSe and SrTiO$_3$, respectively. The two thick spectra are the averaged DCB curve (red) and the averaged SC curve (blue). For these two types of spectra, the former (DCB spectra) shows a much larger gap size and the absence of coherence peaks, which is clearly different from the latter (SC spectra).

### B. Other possibilities as the explanations of V-shaped gap in 1-UC FeSe

We point out that in the regions far from Pb nano-islands, the FeSe film's gap remains U-shaped as reported before, but only in-between two close Pb nano-islands, the FeSe film shows a V-shaped gap instead due to the strengthened scattering potential from the Pb nano-islands as scatterers. Inverse proximity effect (IPE) of Pb nanocrystals on FeSe, if any, will weaken the superconductivity of FeSe, making the gap be filled with quasiparticle excitations and thus turn V-shaped. Based on the following arguments, the possibility of IPE can be excluded. i) The IPE scenario should in principle show no preference and be applicable to both individually isolated and sufficiently close doublet islands, while in experiments the V-gapped spectra only exist in the latter of these two situations. ii) FeSe shows no SC proximity effect on Pb islands, which thus likely exert no IPE on FeSe. iii) The weakened superconductivity of V shape due to IPE normally yields elevated zero-bias conductance (ZBC). However, in our experiments, the ZBC for V-gapped spectra at $d_\Delta < d_\Delta^c$ (3.2 nm) remains nearly zero and presents no obvious difference from that for U-shaped spectra at $d_\Delta > d_\Delta^c$ (Fig. S15). Furthermore, suppressed superconductivity in reduced size by quantum confinement might similarly give the V-shaped gap. Yet, the SC coherence length of 1-UC FeSe is ~2 nm [14], smaller than the critical $d_\Delta^c$ ~3.2 nm where the nodal crossover occurs. The absence of confinement-induced suppression of superconductivity can also be attributed to the nonideal nature of the doublet islands as a quantum-confined structure, particularly along the paths deviating from the center-center direction.

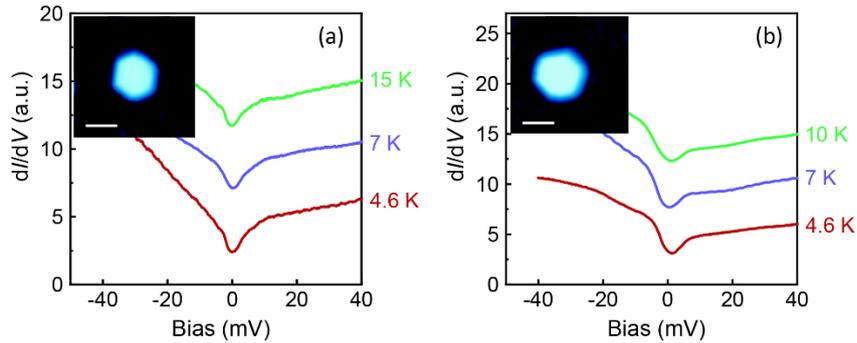

FIG. 12. Temperature dependence of DCB spectra. (a,b) d$I$/d$V$ spectra at different temperatures obtained on smaller-sized Pb islands grown on FeSe. The d$I$/d$V$ curves are shifted vertically for clarity. Scale bar: 5 nm.

## VI. SUMMARY

In summary, nonmagnetic metallic islands grown on a high-$T_c$ superconductor in the Coulomb-blockade regime have been investigated systematically via spectroscopy measurements. We found that the type of Coulomb blockade for tunneling into the nonmagnetic nanoislands may carry information about the Cooper-pairing potential of the high-$T_c$ substrate. For our example of single-crystalline Pb nanostructures grown on 1-UC FeSe, the detected DCB—a direct consequence of the low-impedance contact between the Pb islands and 1-UC FeSe, can arise due to the partly filled tunneling-barrier-type SC gap by the intragap excitations induced by nonmagnetic scatterings from Pb nanoislands. Independent quantum-transport calculations confirm that the transmission for a Pb/FeSe contact with excitation-filled gap indeed falls in the DCB region. Based on the rules of impurity-scattering effects for different pairings [43], these results likely point to the sign-changing gap of the underlying 1-UC FeSe. Further theoretical modelling is required, but is beyond the scope of this work, for a thorough understanding of the detailed interplay between Cooper-pair scatterings by nonmagnetic impurities, interface contacts, and Coulomb blockade.

The conclusion (i.e. sign reversal) as inferred from DCB is independently supported by the measured U–V crossover of the gap line shapes, which reveals a nodal-gap structure, as typically expected for a sign-reversing superconductor after nodal transition. The Coulomb blockade, combined with the induced V-gap transition, consistently offers a route to double-check whether the nodal lines, or sign-reversing pairing, exist in a large-gap U-shaped superconductor. The Coulomb blockade for nonmagnetic islands may be extended on flourishing superconductor substrates and generalized as a generic phase-sensitive method for probing the SC order parameter.

## ACKNOWLEDGMENTS

The authors acknowledge assistance from Linqiang Xu in the quantum-transport calculations. This work was financially supported by National Natural Science Foundation of China (No. 12488201), National Key R&D Program of China (No. 2018YFA0305604), and Innovation Program for Quantum Science and Technology (No. 2021ZD0302403). We also acknowledge support from the Research Council of Finland through the Finnish Centre of Excellence in Quantum Technology (No. 352925).

Supplementary Material for

# Dynamical Coulomb Blockade as a Signature of the Sign-Reversing Cooper Pairing Potential

Chaofei Liu,[1,2] Pedro Portugal,[3] Yi Gao,[4] Jie Yang,[5] Xiuying Zhang,[6] Yanzhao Liu,[1] Tianheng Wei,[1] Wei Ren,[1] Jing Lu,[6,7,8,9] Christian Flindt,[3] and Jian Wang[1,7,10,†]

[1]*International Center for Quantum Materials, School of Physics, Peking University, Beijing 100871, China*
[2]*School of Physics and Wuhan National High Magnetic Field Center, Huazhong University of Science and Technology, Wuhan 430074, China*
[3]*Department of Applied Physics, Aalto University, Aalto 00076, Finland*
[4]*Center for Quantum Transport and Thermal Energy Science, Jiangsu Key Lab on Opto-Electronic Technology, School of Physics and Technology, Nanjing Normal University, Nanjing 210023, China*
[5]*Key Laboratory of Material Physics, Ministry of Education and School of Physics, Zhengzhou University, Zhengzhou 450001, China*
[6]*State Key Laboratory for Mesoscopic Physics and School of Physics, Peking University, Beijing 100871, China*
[7]*Collaborative Innovation Center of Quantum Matter, Beijing 100871, China*
[8]*Beijing Key Laboratory for Magnetoelectric Materials and Devices (BKL-MEMD), Beijing 100871, China*
[9]*Peking University Yangtze Delta Institute of Optoelectronics, Nantong 226000, China*
[10]*Hefei National Laboratory, Hefei 230088, China*

[†]jianwangphysics@pku.edu.cn

# SUPPLEMENTARY FIGURES

## I. STM Characterizations of Pb Nanocrystals on 1-UC FeSe/SrTiO$_3$

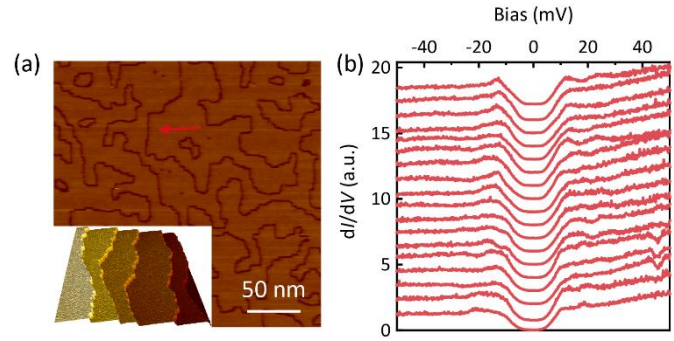

FIG. S1. (a) STM topography of a step-free 1-UC FeSe region. Inset: a large-scale topographic image (500×500 nm$^2$). (b) Spatially resolved tunneling spectra (vertically offset for clarity) taken along the arrow-indicated straight-line trajectory on FeSe surface in (a). As previous reports [1], the spectra show the double-SC-gap feature, characteristic of the multiband superconductivity in 1-UC FeSe. Notably, these spectra taken along the 40-nm-long trajectory are stable in lineshape with negligible change, together with the atomically flat morphology over >500 nm shown in (a), suggesting the high crystal quality of our 1-UC FeSe sample.

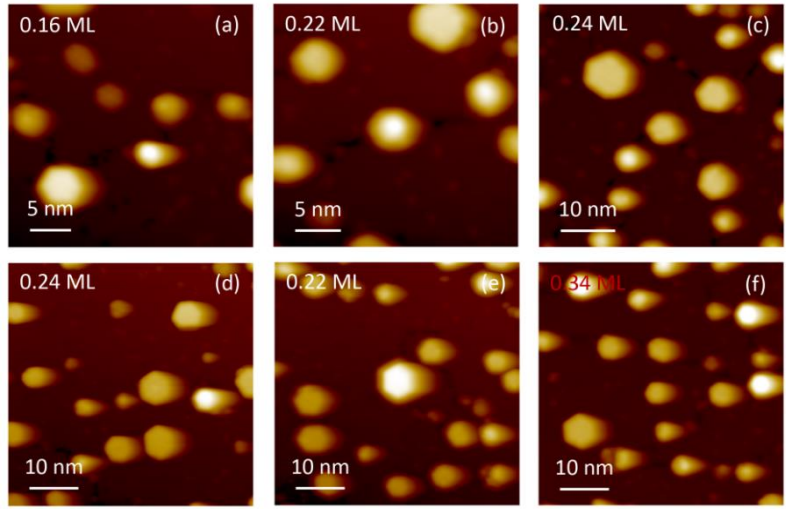

FIG. S2. STM topographic images of Pb nanocrystals with different coverages deposited on 1-UC FeSe. A majority of the Pb nanocrystals reproducibly show well-defined hexagonal shapes with flat top, suggesting their high crystallinity.

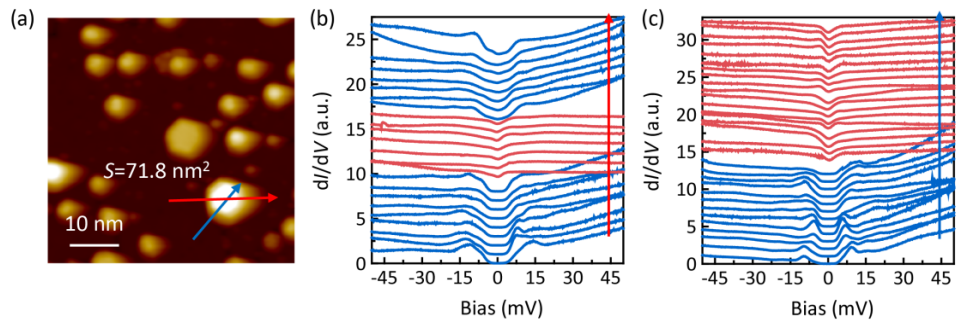

FIG. S3. Spatial homogeneity of the DCB spectra taken on Pb nanocrystals. (a) STM topographic image of Pb nanocrystals on 1-UC FeSe. (b,c) Tunneling spectra (vertically offset) taken along the arrows in (a). The data exemplify nearly unchanged spectral lineshapes of SC gap (blue curves) and DCB gap (red curves) on 1-UC FeSe and Pb nanocrystals, respectively. The uniform d$I$/d$V$ spectra in space collected for a given island justify its high crystallinity, which warrants further analysis based a single spectrum for each island.

## II. DCB Gap for Tunneling into Pb Nanocrystals

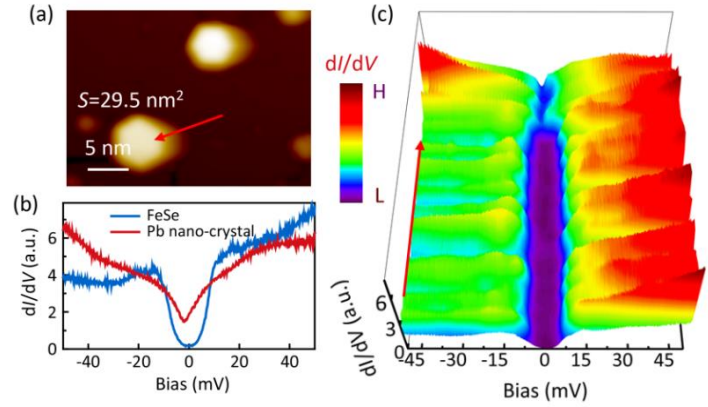

FIG. S4. Additional DCB example #1. (a) STM image of Pb nanocrystals on 1-UC FeSe. (b) Tunneling spectra taken on 1-UC FeSe and Pb nanocrystal, respectively. (c) 3D false-color plot of the tunneling spectra taken along the arrow in (a).

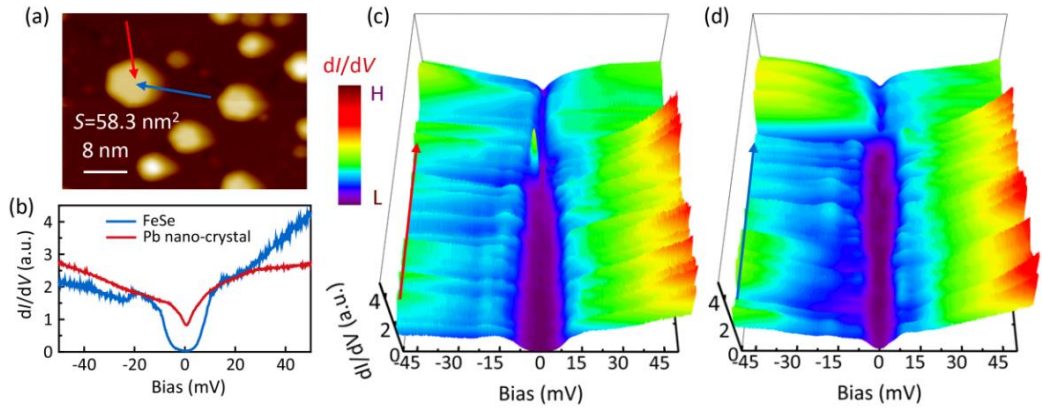

FIG. S5. Additional DCB example #2. (a) STM topographic image of Pb nanocrystals on 1-UC FeSe. (b) Tunneling spectra taken on 1-UC FeSe and Pb nanocrystal, respectively. (c,d) 3D false-color plots of the tunneling spectra taken along the arrows in (a).

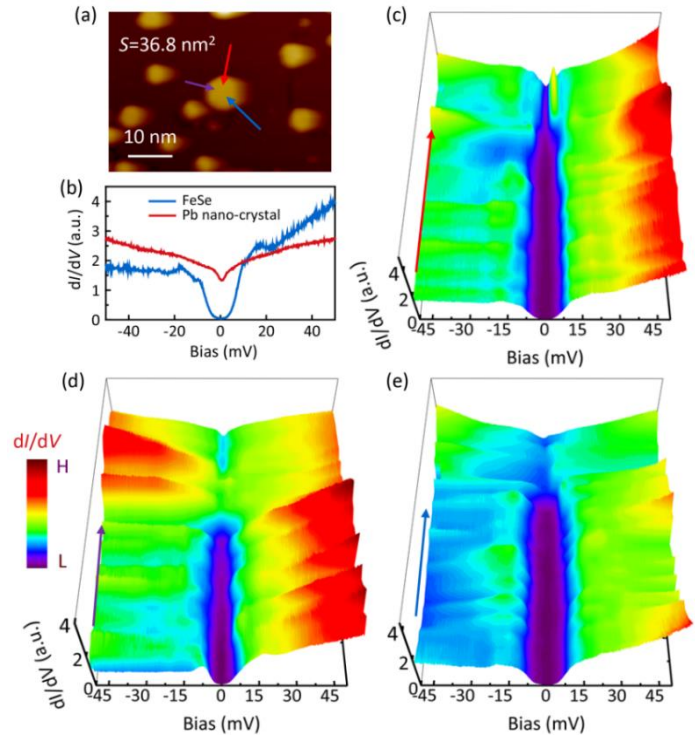

FIG. S6. Additional DCB example #3. (a) STM topographic image of Pb nanocrystals on 1-UC FeSe. (b) Tunneling spectra taken on 1-UC FeSe and Pb nanocrystal, respectively. (c–e) 3D false-color plots of the tunneling spectra taken along the arrows in (a).

## III. Sharp Spectral Transition at the Boundary Between FeSe and Pb Nanocrystal

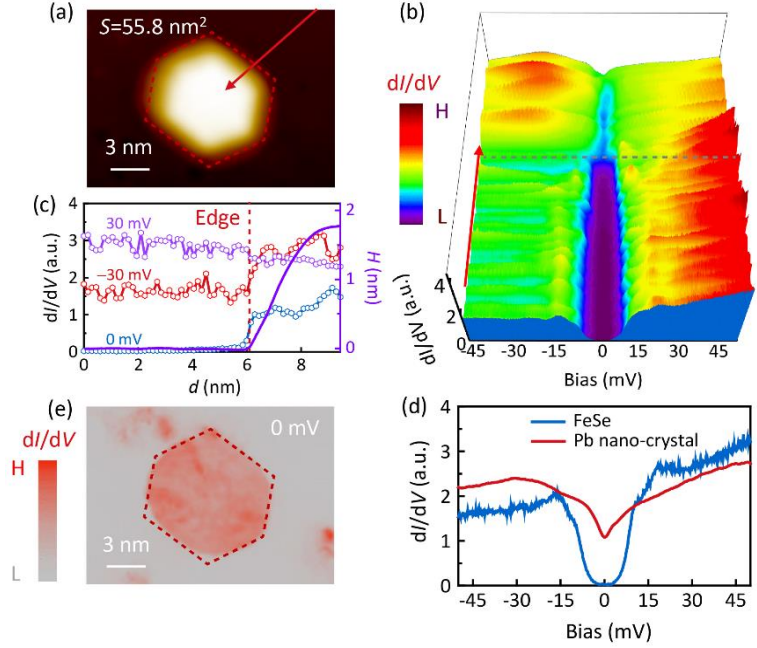

FIG. S7. Additional example of sharp transition of the spectral lineshape across Pb nanocrystal edge. (a) STM topographic image of a Pb nanocrystal. (b) 3D false-color plot of the tunneling spectra taken along the arrow in (a). (c) Differential-conductance d$I$/d$V$(0 mV, ±30 mV) linecuts and height $H$ profile across the Pb nanocrystal edge along the arrow in (a). (d) Tunneling spectra taken on 1-UC FeSe and Pb nanocrystal, showing SC gap and DCB gap, respectively. (e) d$I$/d$V$ map taken at 0 mV for the Pb nanocrystal in (a). The dashed hexagons in (a) and (e) depict the profile of the Pb nanocrystal.

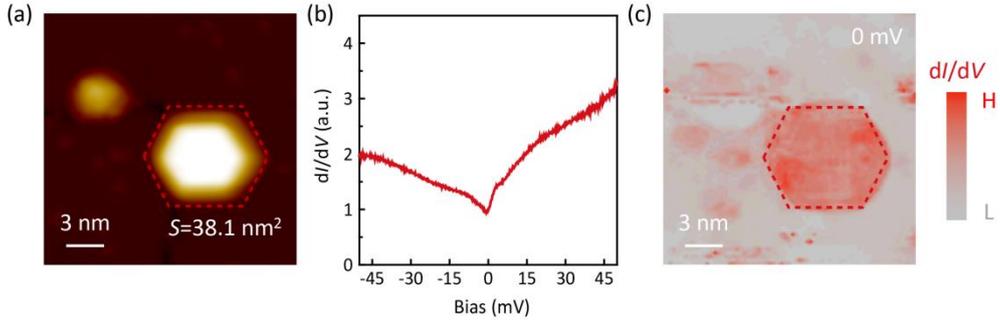

FIG. S8. (a) STM topographic image of a Pb nanocrystal. (b,c) Tunneling spectra and d$I$/d$V$ map taken at 0 mV for the Pb nanocrystal in (a). The well-preserved, nearly size-unchanged hexagonal shape in d$I$/d$V$(0 mV) map for the Pb nanocrystal suggests the sharp spectral transition at nanocrystal edge.

## IV. Control Experiments for Larger-Sized Pb Islands

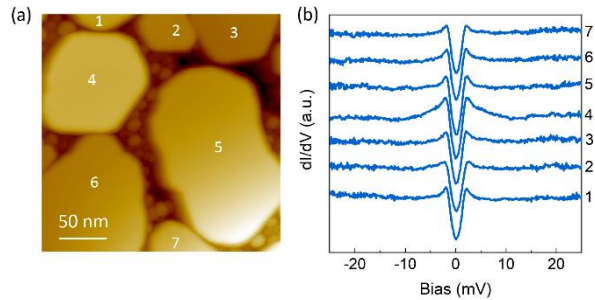

FIG. S9. (a) STM image of larger-sized Pb islands (≳50–100 nm). (b) Tunneling spectra (vertically offset) taken on the numbered islands in (a), showing SC characteristics.

## V. Normalization of the Tunneling Spectra

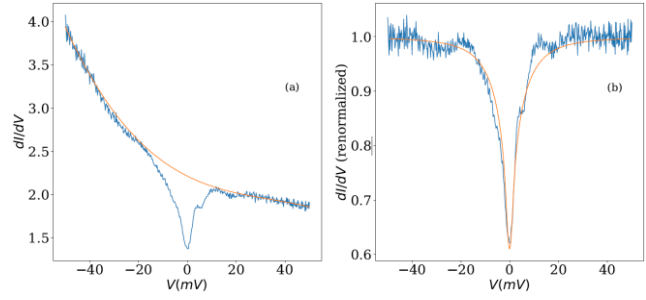

FIG. S10. In (a), we fitted the raw tunneling spectrum (blue) with a third-degree polynomial (yellow) for $|V| > 35$ mV, and divided the raw spectrum by it. Then, in (b), the normalized spectrum (blue) was fitted according to the DCB model (yellow) [2].

## VI. U–V Crossover for SC Lineshape of 1-UC FeSe

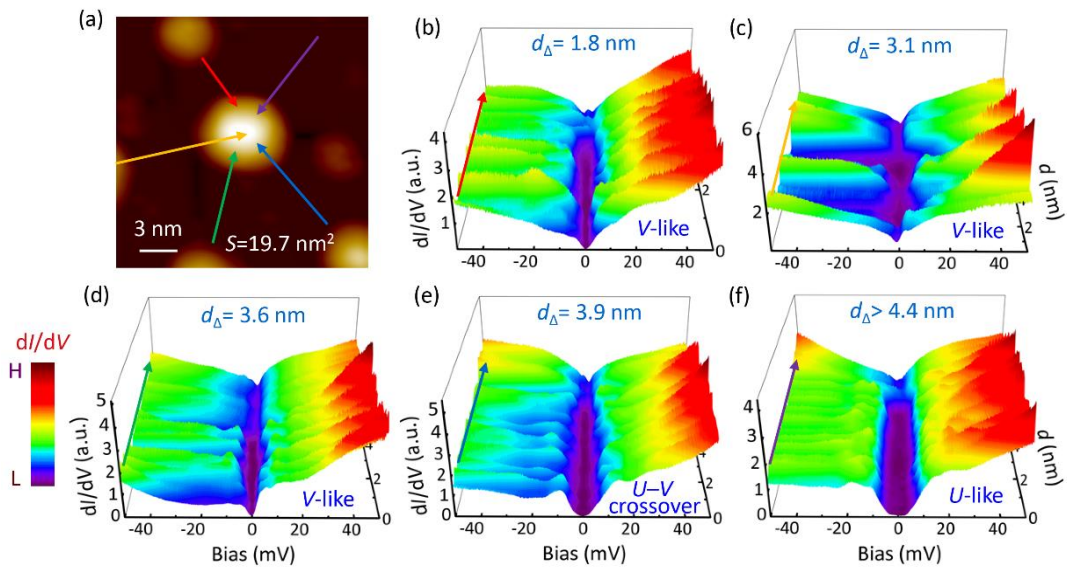

FIG. S11. (a) STM topographic image of Pb nanocrystals. (b–f) 3D false-color plots of the tunneling spectra taken along the arrows in (a). Evidently, as increasing the inter-island distance $2d_\Delta$, U–V crossover occurs for the SC-gap lineshape of 1-UC FeSe.

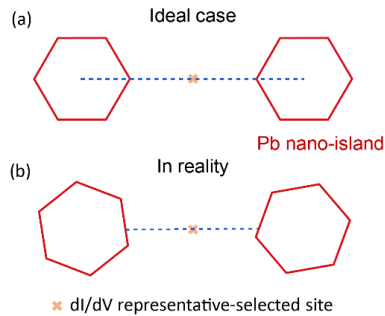

FIG. S12. Schematic showing the 'middle site' of doublet islands. In the ideal case, doublet islands are aligned unidirectionally, while in reality, twisted alignment exists.

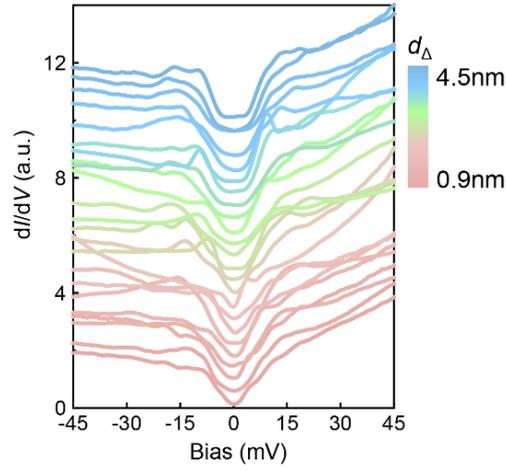

FIG. S13. Raw data (vertically offset) for Fig. 8(e).

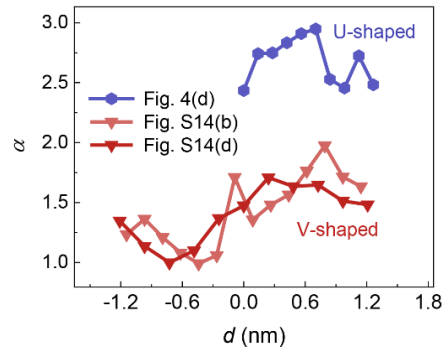

FIG. S14. $\alpha$ vs. $d$, the distance relative to the middle site of doublet islands, extracted from Figs. 8(d), S11(b) and S11(d) with relatively non-uniform spectral variations in space. Note that the isolated island can be taken as an extreme case of the doublet islands with large enough distance between the two islands. Despite fluctuations, $\alpha$ remains sharply different for U- and V-shaped spectra within $d \approx \pm 1$ nm, warranting that the $\alpha$ at the middle site ($d=0$) is representative for the statistical analysis of U–V transition shown in Fig. 8(f).

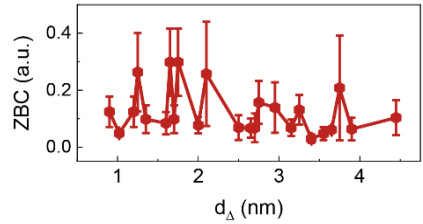

FIG. S15. ZBC as a function of $d_\Delta$. The error bar is determined as half the difference between maximum and minimum ZBC extracted from a set of spectra for a fixed $d_\Delta$.

## VII. DoS Calculations at the Inter-impurity Site under Various Pairing Scenarios

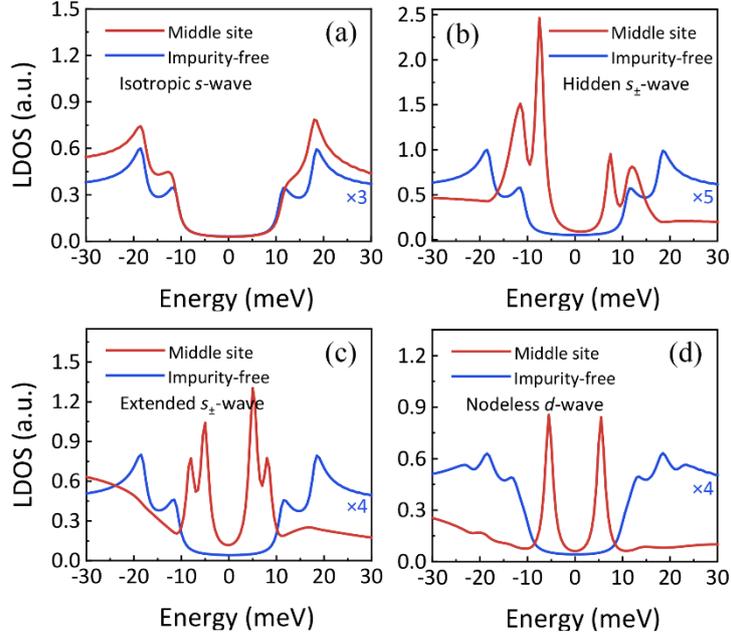

FIG. S16. Spectral calculations under various pairing scenarios. The DoS spectra (red) are calculated upon the middle point of two nonmagnetic impurities in (a) isotropic $s$-, (b) hidden $s_\pm$-, (c) extended $s_\pm$- and (d) nodeless $d$-wave pairing states, respectively. The impurities are placed at the Fe atoms of sublattice A with the location sites of $\boldsymbol{R}_J = (N/2-2, N/2)$ and $\boldsymbol{R}_{J'} = (N/2, N/2)$. The calculated DoS for the impurity-free system in each pairing case (blue) is shown for comparison.